\documentclass[onecolumn,showpacs]{revtex4}

\usepackage{graphicx}
\usepackage{dcolumn}
\usepackage{bm}

\input epsf

\begin{document}
\title{\Large The Effects of Tachyonic and Phantom Fields in the Intermediate and Logamediate Scenarios of the Anisotropic
Universe}

\author{\bf Shuvendu Chakraborty$^1$\footnote{shuvendu.chakraborty@gmail.com } and Ujjal Debnath$^2$\footnote{ujjaldebnath@yahoo.com,
ujjal@iucaa.ernet.in}}

\affiliation{$^1$Department of Mathematics,  Seacom   Engineering   College,  Howrah-   711 302,  India.\\
$^2$Department of Mathematics, Bengal Engineering and Science
University, Shibpur, Howrah-711 103, India.}

\date{\today}

\begin{abstract}
In this work, we have analyzed two scenarios namely,
``intermediate'' and ``logamadiate'' scenarios for closed, open
and flat anisotropic universe in presence of phantom field, normal
tachyonic field and phantom tachyonic field. We have assumed that
there is no interaction between the above mentioned dark energy
and dark matter. In these two types of the scenarios of the
universe, the nature of the scalar fields and corresponding
potentials have been investigated. In intermediate scenario, (i)
the potential for normal tachyonic field decreases, (ii) the
potentials for phantom tachyonic field and phantom field increase
with the corresponding fields. Also in logamediate scenario, (i)
the potential for normal tachyonic field increases, (ii) the
potentials for phantom tachyonic field and phantom field decrease
with the corresponding fields.
\end{abstract}


\maketitle

\section{\normalsize\bf{ INTRODUCTION}}

In recent observations it is strongly believed that the Universe
is experiencing an accelerated expansion. The observation from
type Ia supernovae [1] in associated with Large scale Structure
[2] and Cosmic Microwave Background anisotropies [3] have shown
the evidences to support cosmic acceleration. The main theory
responsible for this scenario is the theory of dark energy. This
mysterious dark energy with negative pressure leads to this cosmic
acceleration. Also the observations indicate that the dominating
component of the present Universe is this dark energy. Dark energy
occupies about 73\% of the energy of our Universe, while dark
matter about 23\% and the usual baryonic matter 4\%. There are
different candidates obey the property of dark energy to violate
the strong energy condition $\rho+3p>0$ given by $-$ quintessence
[4], K-essence [5], Tachyon [6], Phantom [7], ghost condensate
[8,9] and quintom [10], interacting dark energy models [11], brane
world models [12] and Chaplygin gas models [13]. In [14] Chang
etal studied the Phantom field $\phi$ with the potential
$V(\phi)=V_{0}\exp(-\lambda\phi^{2})$ and the dark matter in the
spatially flat FRW model with attractor solutions depending on
$\lambda$. In [15] Shang-Gang Shi et al discussed the cosmological
evaluation of a dark energy model with two scalar fields - Tachyon
and the other Phantom Tachyon where the equation of state $w$
changes from $w>-1$ to $w<-1$ during the evaluation of the
Universe which is a quintom like behavior. In [16] Benaoum has
studied the behaviour of modified Chaplygin gas and effect on the
accelerating Universe in FRW model. In [17] Sami has discussed
cosmological prospect of rolling Tachyon with exponential
potential. In [18] Debnath has shown that the emergent scenario is
possible for the closed Universe if the Universe contains the
normal Tachyon field and for the Phantom field (or Tachyonic
field) the emergent scenario is possible for flat, open and closed
Universe. The holographic description of Tachyon dark
energy in FRW model has been studied by Setare [19].\\

Motivated from the consistency of observational measurement of
CMB about the spectral index and ratio of tensor to scalar
perturbations, we have considered two pre-assigned form of scale
factors (backward approach) as: (i) ``intermediate scenario" and
(ii) ``logamediate scenario" [16, 17] to study of the expanding
anisotropic Universe in the presence of tachyon field and phantom
scalar field. This approach is new as we have studied the
expansion of the universe in anisotropic model, where we consider
the two scale factors independently follow the said scenarios. In
the first case the scale factors evolve separately as
$a(t)=\exp(A^{f_{1}})$ and $b(t)=\exp(B^{f_{2}})$. So the
expansion of the Universe is slower than standard de Sitter
inflation (arises when $f_{1} = f_{2} = 1$) but faster than power
law inflation with power greater than 1. The Harrison - Zeldovich
spectrum of fluctuation arises when $f_{1} = f_{2} = 1$ and
$f_{1} = f_{2} = 2/3$. In the second case we analyze the
inflation with scale factors separately of the form
$a(t)=\exp(A(\ln t)^{\lambda_{1}})$ and $b(t)=\exp(B(\ln
t)^{\lambda _{2}})$. When $\lambda_{1}=\lambda _{2}=1$ this model
reduces to power law inflation. The logamediate inflationary form
is motivated by considering a class of possible cosmological
solutions with indefinite expansion which result from imposing
weak general conditions on the cosmological model.\\

\section{\normalsize\bf{Basic Equations and Solutions}}

We  consider homogeneous and anisotropic  space-time  model
described by the line element

\begin{equation}
ds^{2}=-dt^{2}+a^{2}dx^{2}+b^{2}d\Omega_{k}^{2}
\end{equation}

where  $a$  and  $b$  are scale factors and functions  of  time
$t$ alone : we note that

\begin{eqnarray}d\Omega_{k}^{2}= \left\{\begin{array}{lll}
dy^{2}+dz^{2}, ~~~~~~~~~~~~ \text{when} ~~~k=0 ~~~~ ( \text{Bianchi ~I ~model})\\
d\theta^{2}+sin^{2}\theta d\phi^{2}, ~~~~~ \text{when} ~~~k=+1~~
( \text{Kantowaski-Sachs~ model})\\
d\theta^{2}+sinh^{2}\theta d\phi^{2}, ~~~ \text{when} ~~~k=-1 ~~(
\text{Bianchi~ III~ model})\nonumber
\end{array}\right.
\end{eqnarray}

\quad Here  $k$  is  the  curvature  index  of  the  corresponding
2-space, so  that  the  above  three  types  are  described  by
Thorne [22]  as  flat, closed  and  open respectively.\\

Now we consider the Hubble parameter $H$ and the deceleration
parameter $q$ in terms of scale factor as

\begin{equation}
H=\frac{1}{3}\left(\frac{\dot{a}}{a}+2\frac{\dot{b}}{b}\right)~~
\text{and} ~~ q=-1-\frac{\dot{H}}{H^{2}}
\end{equation}

We consider that the Universe contains normal matter and Phantom
field (or Tachyonic field). The Einstein field equations for the
space time given by the equation (1) are

\begin{equation}
\frac{\ddot{a}}{a}+2\frac{\ddot{b}}{b}=-\frac{1}{2}
(\rho_{\phi}+\rho_{m}+3p_{\phi}+3p_{m})
\end{equation}
and
\begin{equation}
\frac{\dot{b}^{2}}{b^{2}}
+2\frac{\dot{a}}{a}\frac{\dot{b}}{b}+\frac{k^{2}}{b^{2}}=(\rho_{\phi}+\rho_{m})
 \
\end{equation}

where $\rho_{m}$ and $p_{m}$ are the energy density and pressure
of the normal matter with the equation of state given by
$p_{m}=w\rho_{m}$, $-1\leq w\leq 1$ and $\rho_{\phi}$ and
$p_{\phi}$ are the energy density and pressure due to the Phantom
field (or Tachyonic field).\\

Now considering that there do not exist any interaction between
normal matter and the Phantom field (or Tachyonic field), that is
they are separately conserved, the energy conservation equation
for normal matter and the Phantom field (or Tachyonic field) are

\begin{equation}
\dot{\rho}_{m}+3H(p_{m}+\rho_{m})=0
\end{equation}
and
\begin{equation}
\dot{\rho}_{\phi}+3H(p_{\phi}+\rho_{\phi})=0
\end{equation}

From equation (5) we have the expression for energy density of
matter as

\begin{equation}
\rho_{m}=\rho_{0}\left(ab^{2}\right)^{-(w+1)}
\end{equation}

where $\rho_{0}$ is the integration constant.\\\

$\bullet$ \textbf{Tachyonic  field:} The energy density
$\rho_{\phi}$ and pressure $p_{\phi}$ due to the Tachyonic field
field $\phi$ is given by

\begin{equation}
\rho_{\phi}=\frac{V(\phi)}{\sqrt{1-\epsilon \dot{\phi}^{2}}}
\end{equation}

\begin{equation}
p_{\phi}=-V(\phi)\sqrt{1-\epsilon \dot{\phi}^{2}}
\end{equation}

where $V(\phi)$ is the relevant potential for the Tachyonic field
$\phi$. It can be seen that
$\frac{p_{\phi}}{\rho_{\phi}}=-1+\epsilon \dot{\phi}^{2}>-1
~\text{or}~ <-1$ according to normal Tachyon $(\epsilon=+1)$ or
Phantom
Tachyon $(\epsilon=-1)$.\\

From the field equations (3), (4), (8) and (9) the expression for
$\dot{\phi}^{2}$ and $V(\phi)$ are given by

\begin{equation}
\dot{\phi}^{2}=\frac{-\frac{2}{3}(\frac{\ddot{a}}{a}+2\frac{\ddot{b}}{b}
-\frac{\dot{b}^{2}}{b^{2}}
-2\frac{\dot{a}}{a}\frac{\dot{b}}{b})+(w+1)\rho_{m}+\frac{2k}{3b^{2}}}{\epsilon
\rho_{\phi} }
\end{equation}
and
\begin{equation}
(V(\phi))^{2}=-\rho_{\phi}p_{\phi}
\end{equation}

$\bullet$ \textbf{Phantom field:} The energy density and pressure
of the Phantom field $\phi$ are respectively given by

\begin{equation}
\rho_{\phi}=-\frac{1}{2}\dot{\phi}^{2}+V(\phi)
\end{equation}
and
\begin{equation}
p_{\phi}=-\frac{1}{2}\dot{\phi}^{2}-V(\phi)
\end{equation}

where $V(\phi)$ is the Phantom field potential. From the equations
(3), (4), (12) and (13) we have

\begin{equation}
\dot{\phi}^{2}=\frac{2}{3}\left(\frac{\ddot{a}}{a}+2\frac{\ddot{b}}{b}
-\frac{\dot{b}^{2}}{b^{2}}
-2\frac{\dot{a}}{a}\frac{\dot{b}}{b}\right)+(w+1)\rho_{m}-\frac{2k}{3b^{2}}
\end{equation}
and
\begin{equation}
V(\phi)=\dot{H}+3H^{2}+\frac{1}{2}(w-1)\rho_{m}+\frac{2k}{3b^{2}}
\end{equation}

\begin{figure}
\includegraphics[scale=0.7]{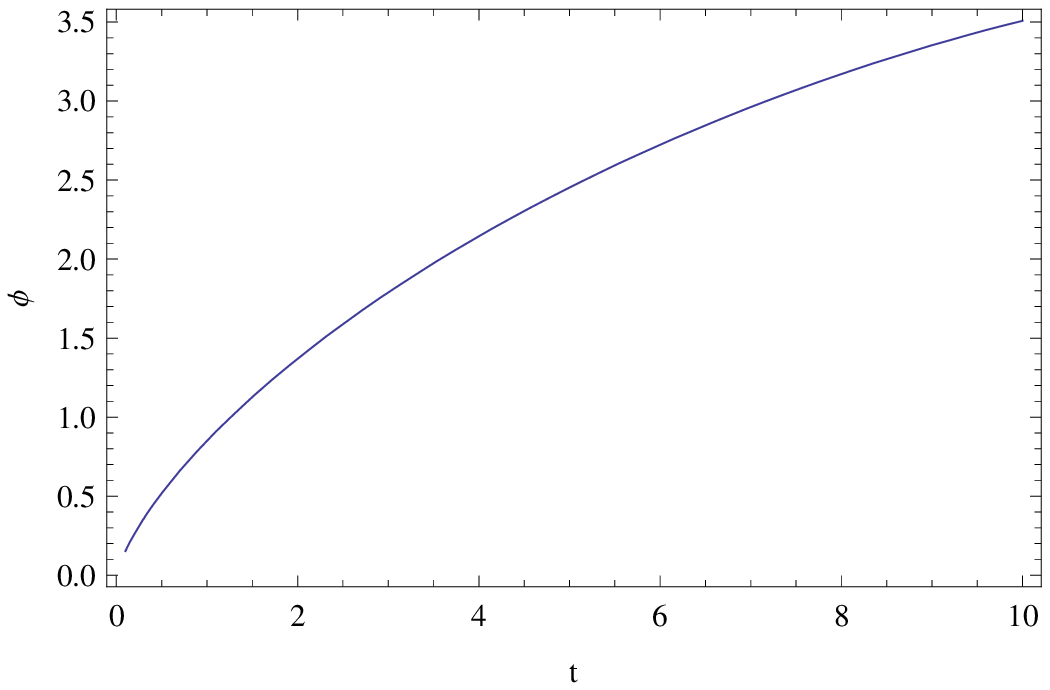}~~~~
\includegraphics[scale=0.7]{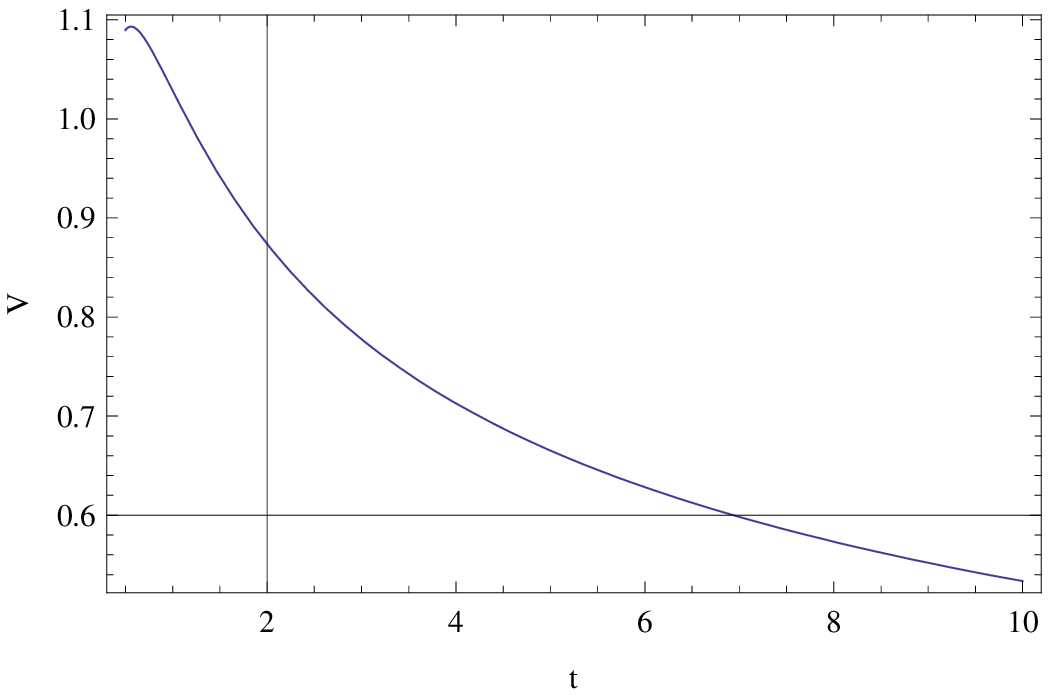}\\
\vspace{2mm}
~~~~~~~Fig.1~~~~~~~~~~~~~~~~~~~~~~~~~~~~~~~~~~~~~~~~~~~~~~~~~~~~~~~~~~~~Fig.2\\
\vspace{6mm}

\includegraphics[scale=0.7]{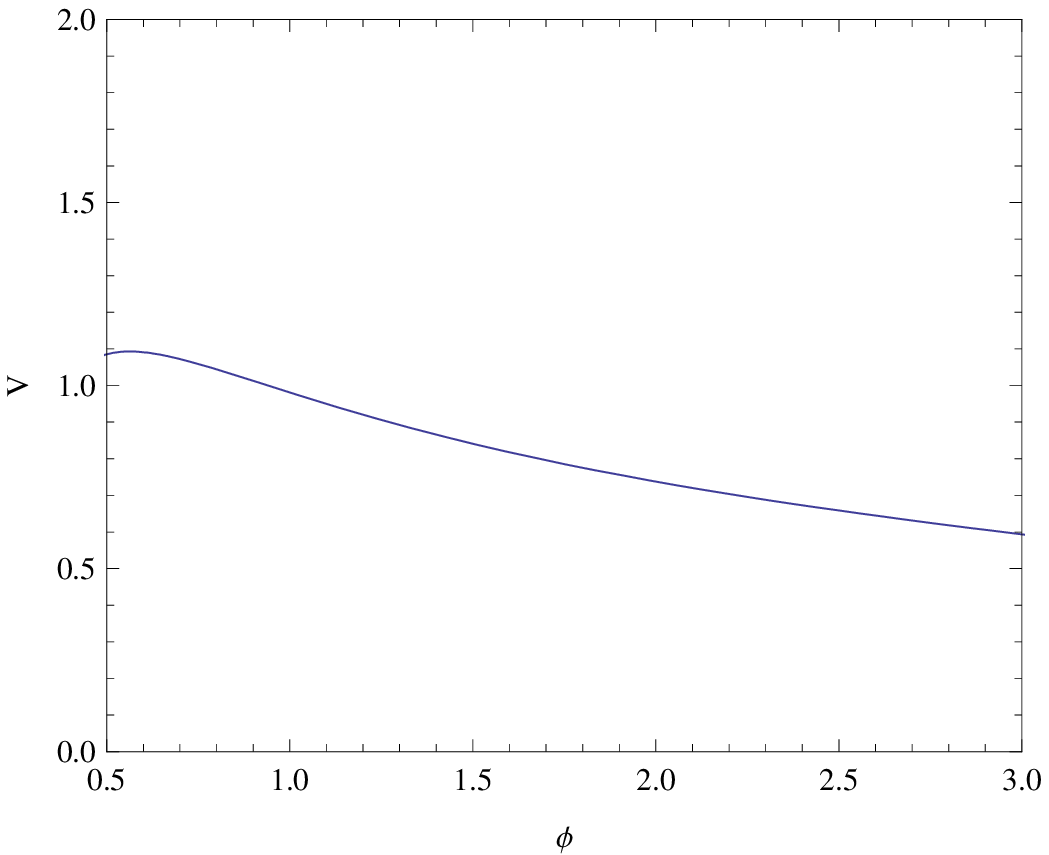}
\vspace{2mm}

Fig.3

\vspace{6mm} Figs. 1 - 2 show the variations of  $\phi$ and $V$
against $t$ and fig. 3  shows the variations of $V$ against
$\phi$, for $A = 1.2, B = 1.1, f_{1} = .7, f_{2} = .6, k = 1, w =
1/3, \rho_{0}= 1$ in presence of normal tachyonic field
($\epsilon=+1$) in intermediate scenario.
 \vspace{6mm}
\end{figure}

\begin{figure}
\includegraphics[scale=0.7]{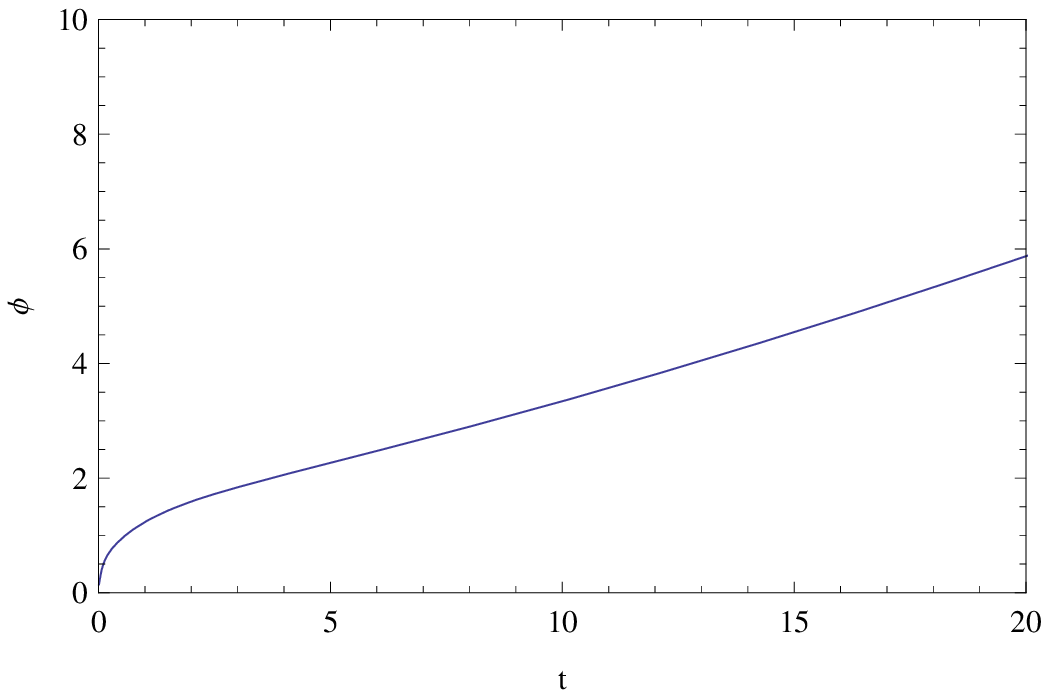}~~~~
\includegraphics[scale=0.7]{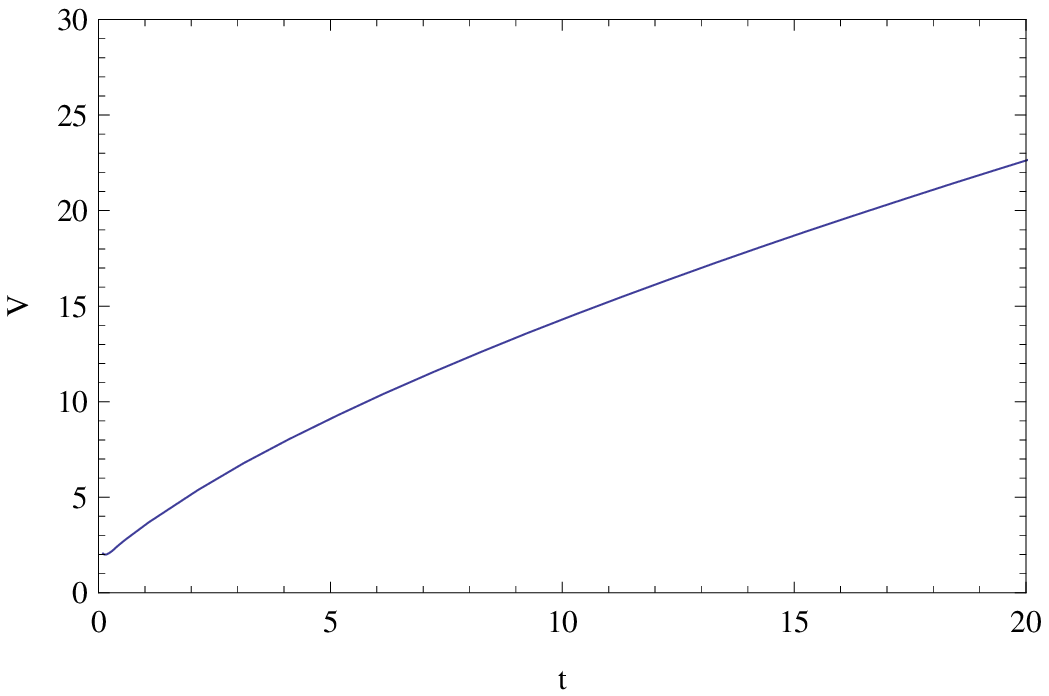}\\
\vspace{2mm}
~~~~~~~Fig.4~~~~~~~~~~~~~~~~~~~~~~~~~~~~~~~~~~~~~~~~~~~~~~~~~~~~~~~~~~~~Fig.5\\
\vspace{6mm}
\includegraphics[scale=0.7]{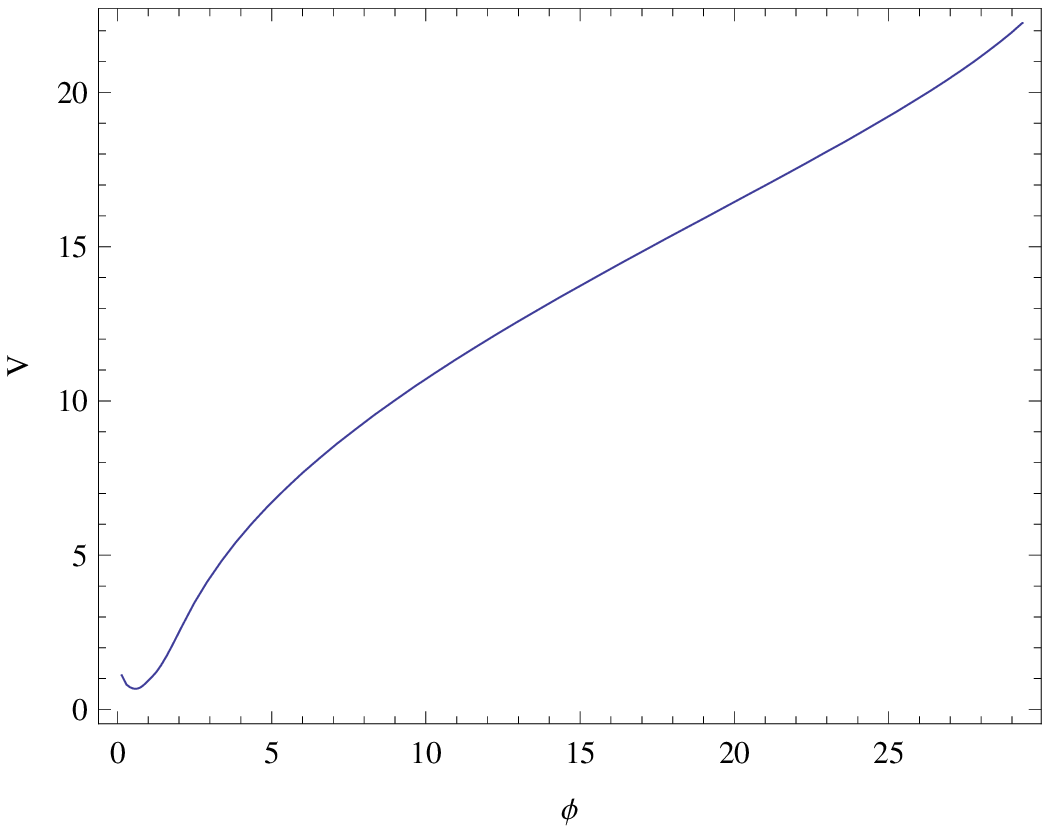}
\vspace{2mm}

Fig.6

\vspace{6mm} Figs. 4 - 5 show the variations of  $\phi$ and $V$
against $t$ and fig. 6  shows the variations of $V$ against
$\phi$, for $A = 1.2, B = 1.1, f_{1} = 1.5, f_{2} = 1.7, k = 1, w
= 1/3, \rho_{0}= 1$ in presence of phantom tachyonic field
($\epsilon=-1$) in intermediate scenario.
\vspace{6mm}
\end{figure}

\begin{figure}
\includegraphics[scale=0.7]{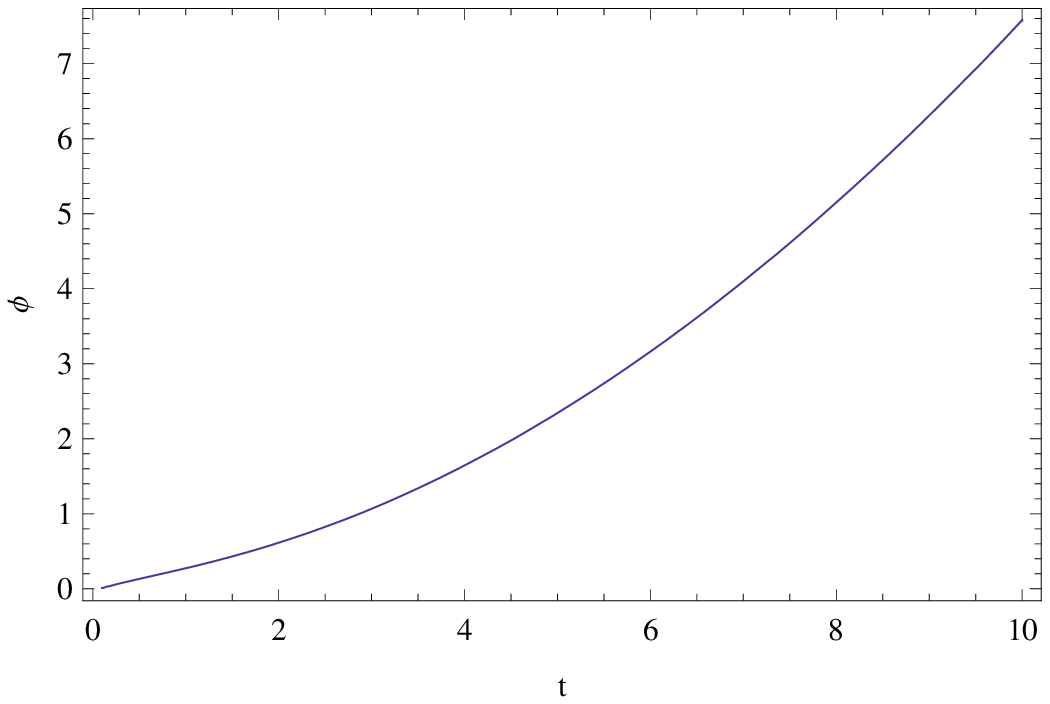}~~~~
\includegraphics[scale=0.7]{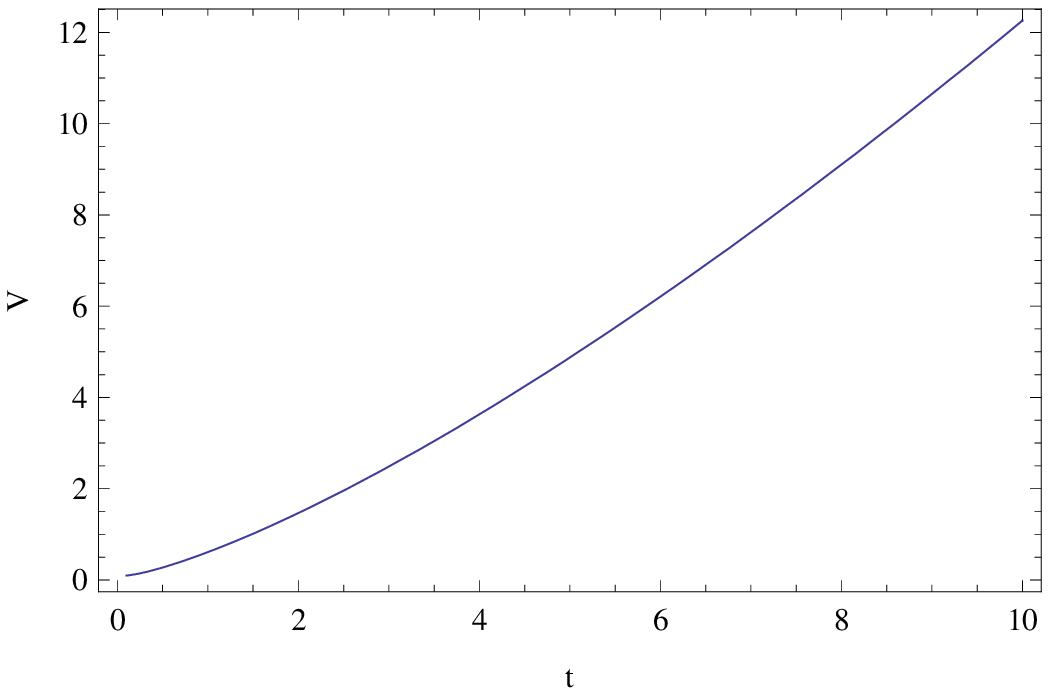}\\
\vspace{2mm}
~~~~~~~Fig.7~~~~~~~~~~~~~~~~~~~~~~~~~~~~~~~~~~~~~~~~~~~~~~~~~~~~~~~~~~~~Fig.8\\
\vspace{6mm}
\includegraphics[scale=0.7]{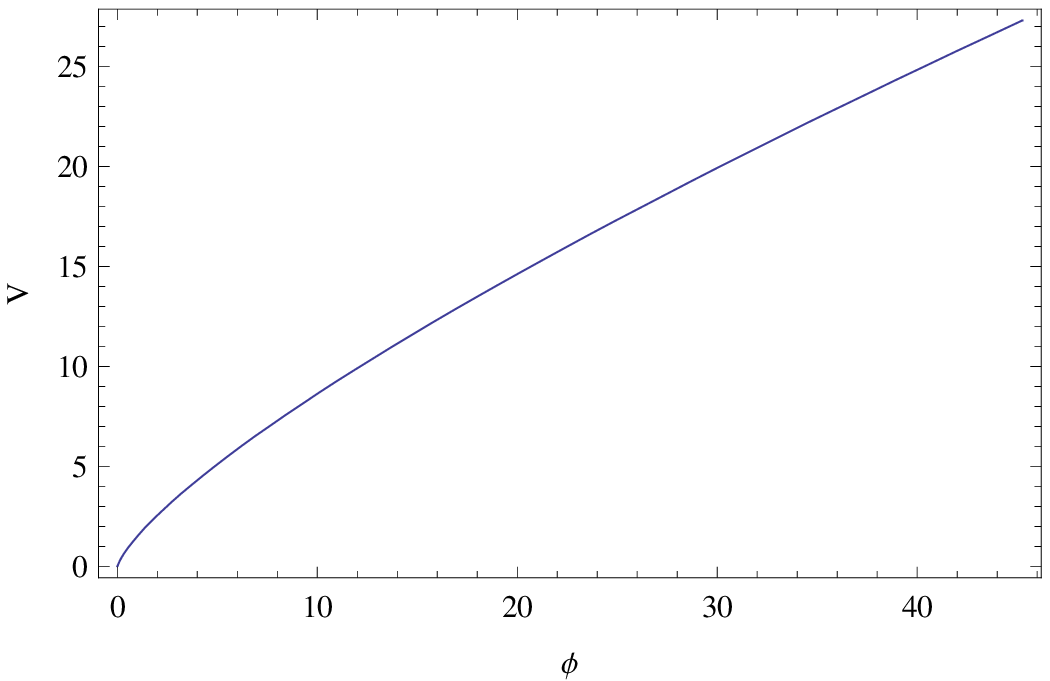}
\vspace{2mm}

Fig.9

\vspace{6mm} Figs. 7 - 8 show the variations of  $\phi$ and $V$
against $t$ and fig. 9  shows the variations of $V$ against
$\phi$, for $A = 2, B = 3, f_{1} = 1.5, f_{2} = 1.7, k = 1, w =
1/3, \rho_{0}= 1$ in presence of phantom field in intermediate
scenario.
\vspace{6mm}
\end{figure}

\subsection{\bf\large{Intermediate  Scenario}}

Here we consider a particular form of intermediate scenario, where
the form of the scale factors $a(t)$ and $b(t)$ are defined as,
[23]

\begin{equation}
a(t)=\exp (A t^{f_{1}})  \quad and  \quad  b(t)=\exp (B t^{f_{2}})
\end{equation}

Using these expressions we have the Hubble parameter and it's
derivatives as,

\begin{equation}
H=\frac{A  t^{f_{1}} f_{1} + 2 B  t^{f_{2}} f_{2}}{3t}, \quad
\dot{H}=\frac{A  t^{f_{1}} f_{1} (- 1 + f_{1}) + 2 B
 t^{f_{2}} f_{2} (- 1 + f_{2})}{3 t^{2}}
\end{equation}

and

\begin{equation}
\ddot{H}=\frac{A t^{f_{1}} f_{1}( 2 + (- 3 + f_{1})  f_{1}) + 2 B
t^{f_{2}} f_{2}( 2 + (- 3 + f_{2})  f_{2})}{3 t^{3}}
\end{equation}

Hence, in presence of normal tachyonic field and for expanding
Universe, $A f_{1}
> 0$ and $B f_{2}
> 0$. Also, from the derivative of Hubble parameter we have $ 0<
f_{1} < 1 $ and  $ 0< f_{2} < 1$ and  then $A > 0$ and $B > 0$.\\

And in presence of phantom field and for a expanding Universe $A
f_{1}
> 0$ and $B f_{2}
> 0$. Also, from the derivative of Hubble parameter we have $
f_{1}
>1 $ and  $  f_{2} > 1$ and  then $A > 0$ and $B > 0$.\\

Now putting the assumed values of the scale factors in (10) and
(11), we have the required expressions for tachyonic field $\phi$
and its potential $V$ as

\begin{equation}
\phi=\int \left[\frac{\frac{2}{3}ke^{-2B
t^{f_{2}}}-\frac{2(At^{f_{1}}f_{1}(-1+(1+At^{f_{1}})f_{1})+B^{2}
t^{2f_{2}}f_{2}^{2}+2Bt^{f_{2}}f_{2}(-1-A t^{f_{1}}f_{1}+f_{2}))}
{3t^{2}}+(e^{A t^{f_{1}}+2Bt^{f_{2}}})^{(-1-w)}(1+w)\rho_{0}}
{\epsilon(e^{-2B t^{f_{2}}}k+2 A B
t^{-2+f_{1}+f_{2}}f_{1}f_{2}+B^{2}t^{-2+2 f_{2}}f_{2}^{2}-(e^{A
t^{f_{1}}+2Bt^{f_{2}}})^{(-1-w)}\rho_{0})}\right]^{1/2}dt
\end{equation}
and
\begin{eqnarray*}
V= \left[e^{-2B t^{f_{2}}}k +2A B
t^{-2+f_{1}+f_{2}}f_{1}f_{2}+B^{2}t^{-2+2 f_{2}}f_{2}^{2}-(e^{A
t^{f_{1}}+2B t^{f_{2}}})^{-1-\omega}\rho_{0}\right]\times
\end{eqnarray*}
\begin{equation}
\left[\frac{\frac{2}{3}ke^{-2B
t^{f_{2}}}-\frac{2(At^{f_{1}}f_{1}(-1+(1+At^{f_{1}})f_{1})+B^{2}
t^{2f_{2}}f_{2}^{2}+2Bt^{f_{2}}f_{2}(-1-A t^{f_{1}}f_{1}+f_{2}))}
{3t^{2}}+(e^{A t^{f_{1}}+2Bt^{f_{2}}})^{(-1-w)}(1+w)\rho_{0}}
{(e^{-2B t^{f_{2}}}k+2 A B
t^{-2+f_{1}+f_{2}}f_{1}f_{2}+B^{2}t^{-2+2 f_{2}}f_{2}^{2}-(e^{A
t^{f_{1}}+2Bt^{f_{2}}})^{(-1-w)}\rho_{0})}\right]^{1/2}
\end{equation}

Now putting the assumed values of the scale factors in (14) and
(15), we have the required expressions for phantom field $\phi$
and its potential V as

\begin{eqnarray*}
\phi=\int \left[\frac{2\left(A
t^{f_{1}}f_{1}(-1+(1+t^{f_{1}})f_{1})+B^{2}t^{2f_{2}}f_{2}^{2}+2Bt^{f_{2}}f_{2}(-1-At^{f_{1}}f_{1}+f_{2})\right)}{3t^{2}}\right.
\end{eqnarray*}
\begin{equation}
\left. + (e^{At^{f_{1}}+2Bt^{f_{2}}})^{-1-w} (1+w)\rho_{0}
-\frac{2}{3}e^{-2Bt^{f_{2}}}k \right]^{1/2} dt
\end{equation}
and
\begin{eqnarray*}
V=\frac{(A t^{f_{1}} f_{1}+2B
t^{f_{2}}f_{2})^{2}+At^{f_{1}}(-1+f_{1})f_{1}+2B
t^{f_{2}}(-1+f_{2})f_{2}}{3 t^{2}}
\end{eqnarray*}
\begin{equation}
+\frac{1}{2}(e^{At^{f_{1}}+2Bt^{f_{2}}})^{-1-w}
(-1+w)\rho_{0}+\frac{2}{3}e^{-2Bt^{f_{2}}}k
\end{equation}

And the dark energy density and mass for of the universe or the
two cases are given by
\begin{equation}
\rho_{\phi}=e^{-2B t^{f_{2}}}k +2A B
t^{-2+f_{1}+f_{2}}f_{1}f_{2}+B^{2}t^{-2+2 f_{2}}f_{2}^{2}-(e^{A
t^{f_{1}}+2B t^{f_{2}}})^{-1-\omega}\rho_{0}
\end{equation}

\begin{equation}
Mass=\frac{e^{A t^{f_{1}}}\left(k t^{2}+B e^{2 B
t^{f_{2}}}t^{f_{2}}f_{2}(2A
t^{f_{1}}f_{1}+Bt^{f_{2}}f_{2})\right)}{t^{2}}
\end{equation}

From above we see that the expressions of tachyonic field and
phantom field and their corresponding potentials are very
complicated. The normal tachyonic field ($\epsilon=+1$) and
corresponding potential against time $t$ have been drawn in
figures 1, 2 respectively and the potential against the
corresponding field have been drawn in figures 3 in intermediate
scenario for $A = 1.2, B = 1.1, f_{1} = .7, f_{2} = .6, k = 1, w =
1/3, \rho_{0}= 1$. Also the phantom tachyon field ($\epsilon=-1$)
and phantom field with the corresponding potentials have been
drawn in figures 4, 5, 7 and 8 respectively and the fields
potentials against the corresponding fields have been drawn in
figures 6 and 9 respectively in intermediate scenario for $A =
1.2, B = 1.1, f_{1} = 1.5, f_{2} = 1.7, k = 1, w = 1/3,
\rho_{0}=1$. From figures 1-3, we see that the normal tachyonic
field always increases with time, potential decreases against time
and normal tachyonic field. From figures 4-6, we see that the
phantom tachyonic field and potential always increase with time
and potential increases with phantom tachyonic field. Also, from
figures 7-9, we see that the phantom field and potential always
increase with time and potential increases with phantom field.\\

\begin{figure}
\includegraphics[scale=0.7]{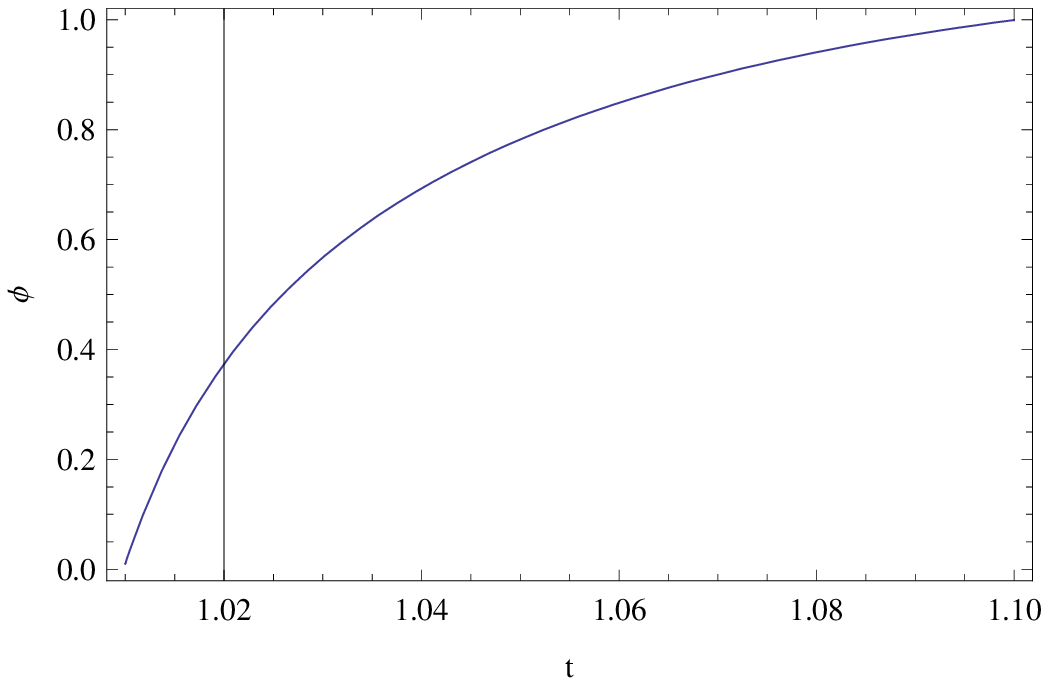}~~~~
\includegraphics[scale=0.7]{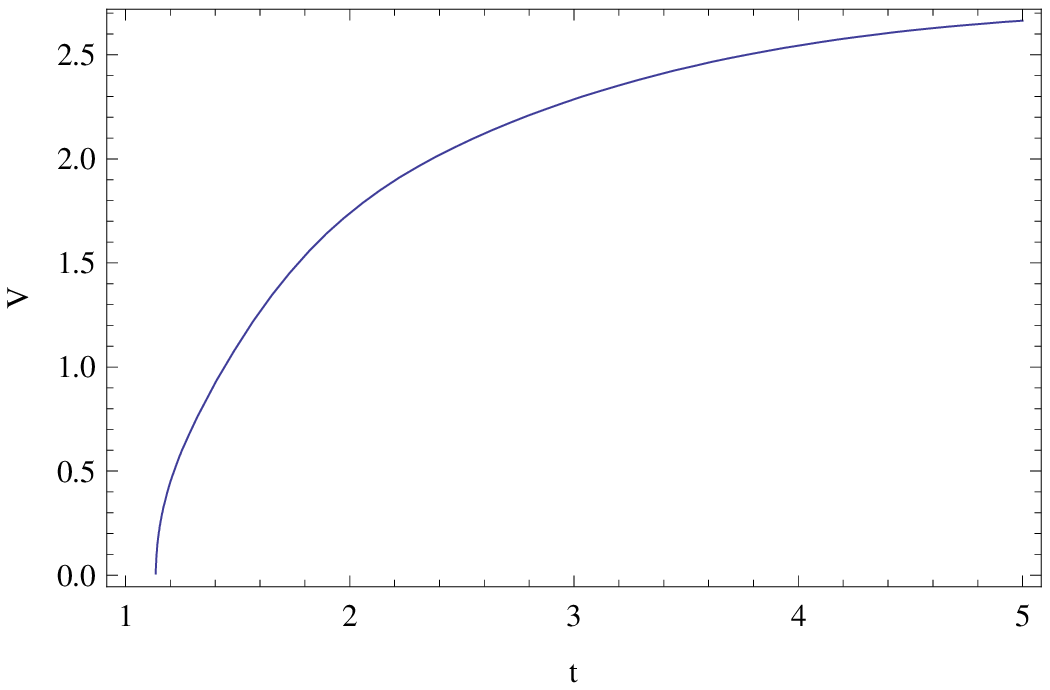}\\
\vspace{2mm}
~~~~~~~Fig.10~~~~~~~~~~~~~~~~~~~~~~~~~~~~~~~~~~~~~~~~~~~~~~~~~~~~~~~~~~~~Fig.11\\
\vspace{6mm}
\includegraphics[scale=0.7]{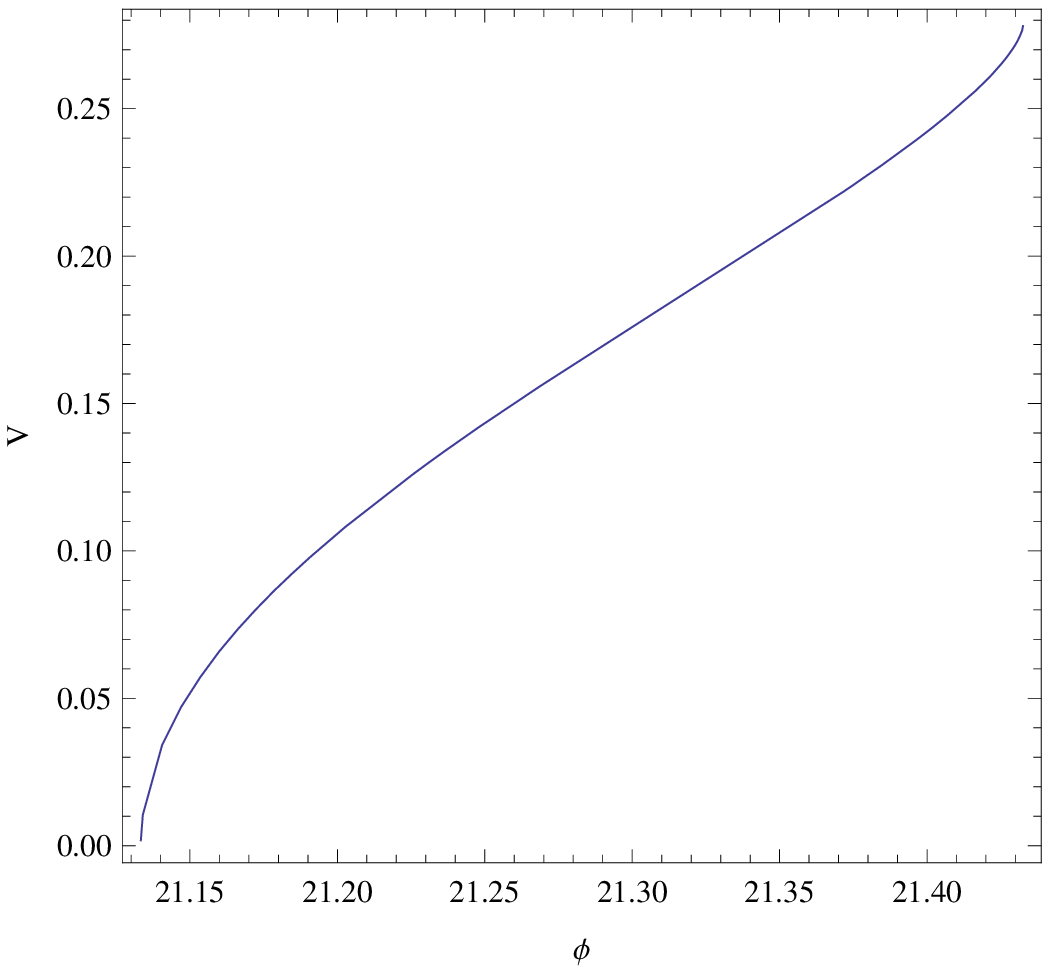}
\vspace{2mm}

Fig.12

\vspace{6mm} Figs. 10 - 11 show the variations
of  $\phi$ and $V$ against $t$ and fig. 11  shows the variations
of $V$ against $\phi$, for $A = 1.1, B = 1.2, \lambda_{1} = 2,
\lambda_{2} = 3, k = 1, w = 1/3, \rho_{0}= 1$ in presence of
normal tachyonic field ($\epsilon=+1$) in logamediate scenario.
 \vspace{6mm}
\end{figure}

\begin{figure}
\includegraphics[scale=0.7]{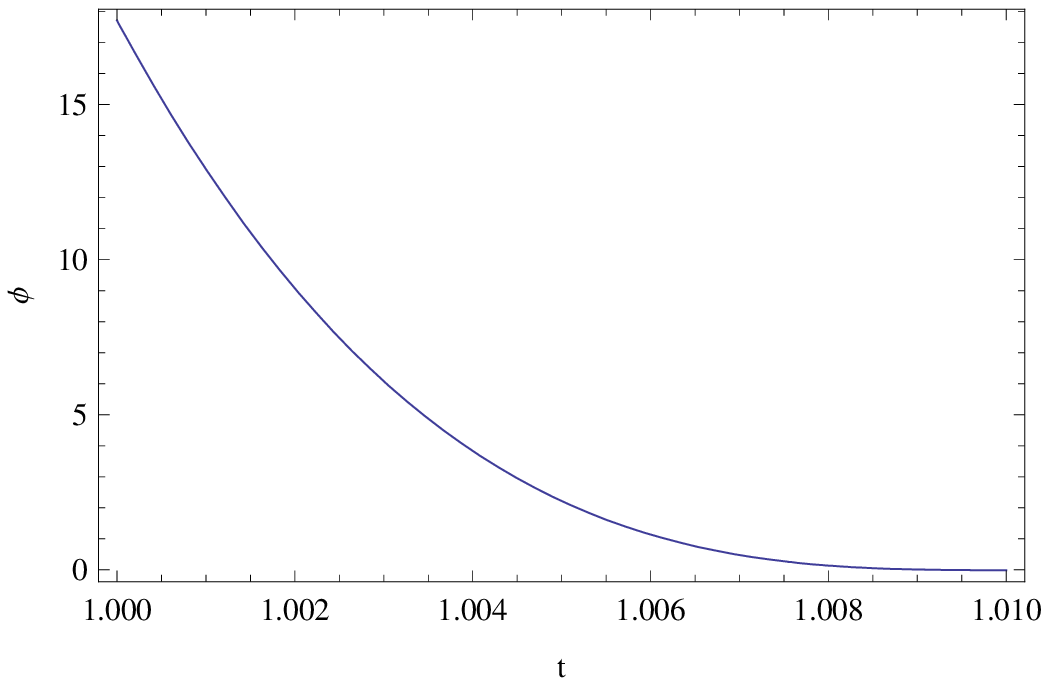}~~~~
\includegraphics[scale=0.7]{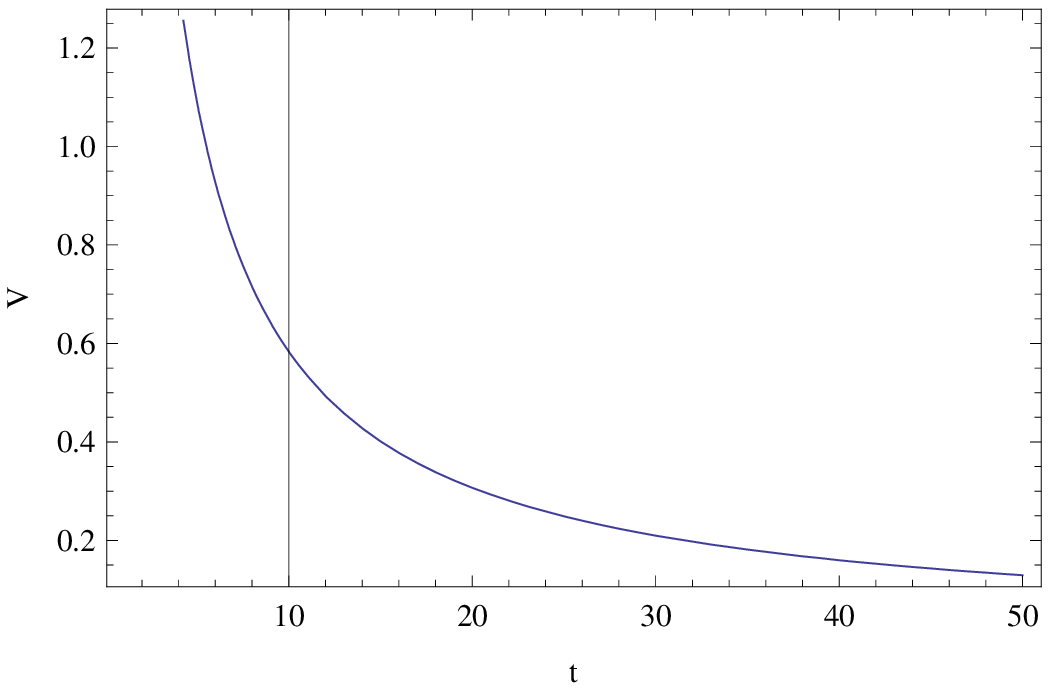}\\
\vspace{2mm}
~~~~~~~Fig.13~~~~~~~~~~~~~~~~~~~~~~~~~~~~~~~~~~~~~~~~~~~~~~~~~~~~~~~~~~~~Fig.14\\
\vspace{6mm}
\includegraphics[scale=0.7]{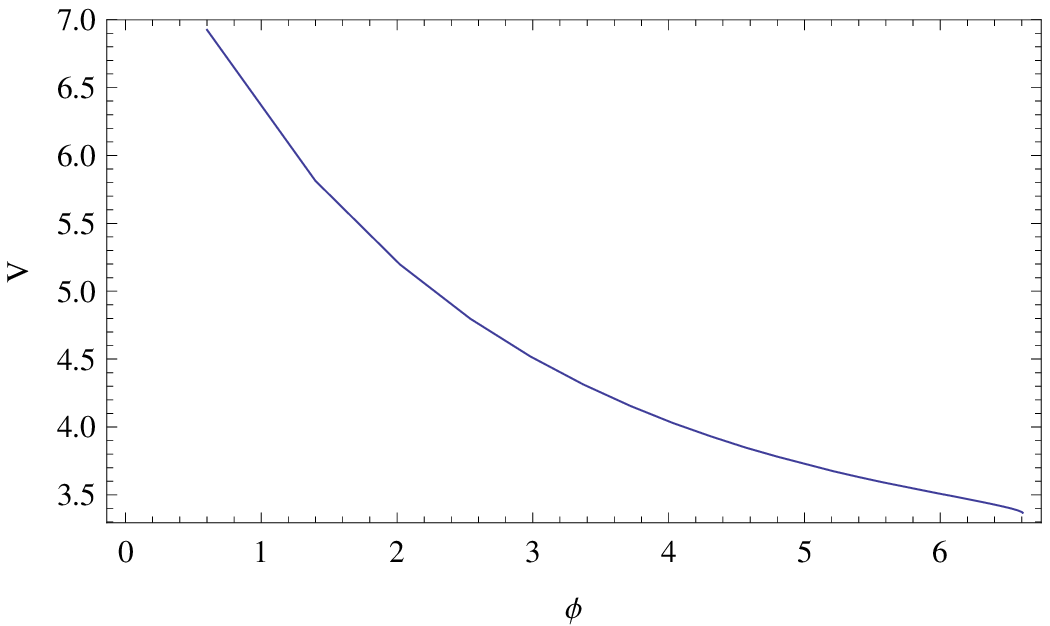}
\vspace{2mm}

Fig.15

\vspace{6mm} Figs. 13 - 14 show the variations of  $\phi$ and $V$
against $t$ and fig. 15  shows the variations of $V$ against
$\phi$, for $A = 1.1, B = 1.2, \lambda_{1} = 2, \lambda_{2} = 3, k
= 1, w = 1/3, \rho_{0}= 1$ in presence of phantom tachyonic field
($\epsilon=-1$) in logamediate scenario.
\vspace{6mm}
\end{figure}

\begin{figure}
\includegraphics[scale=0.7]{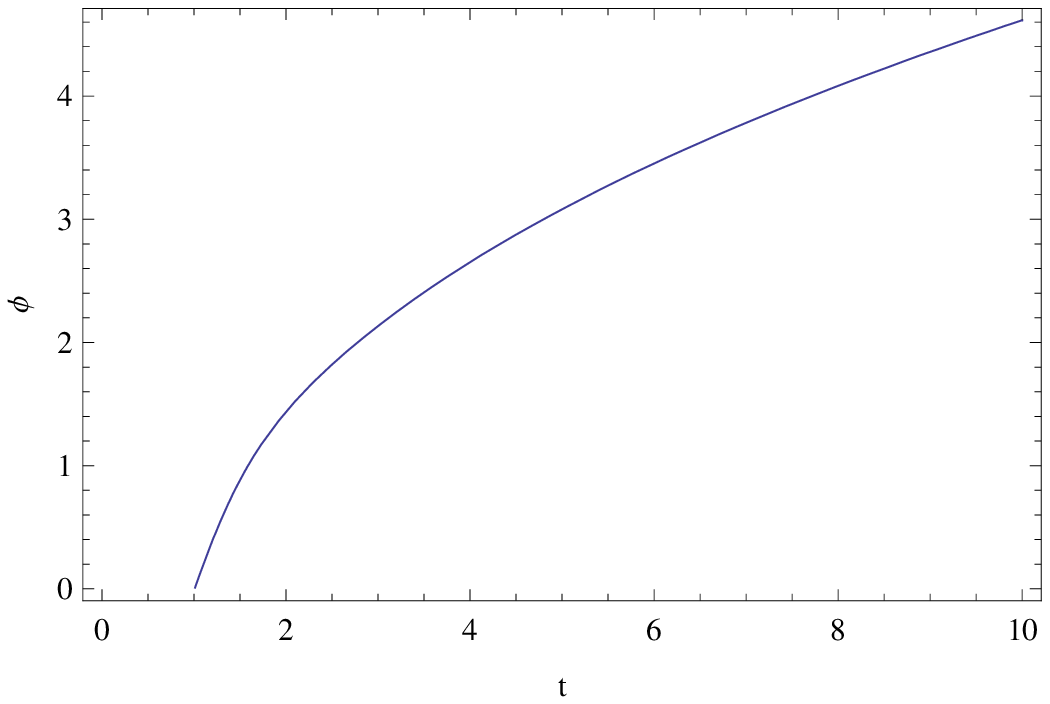}~~~~
\includegraphics[scale=0.7]{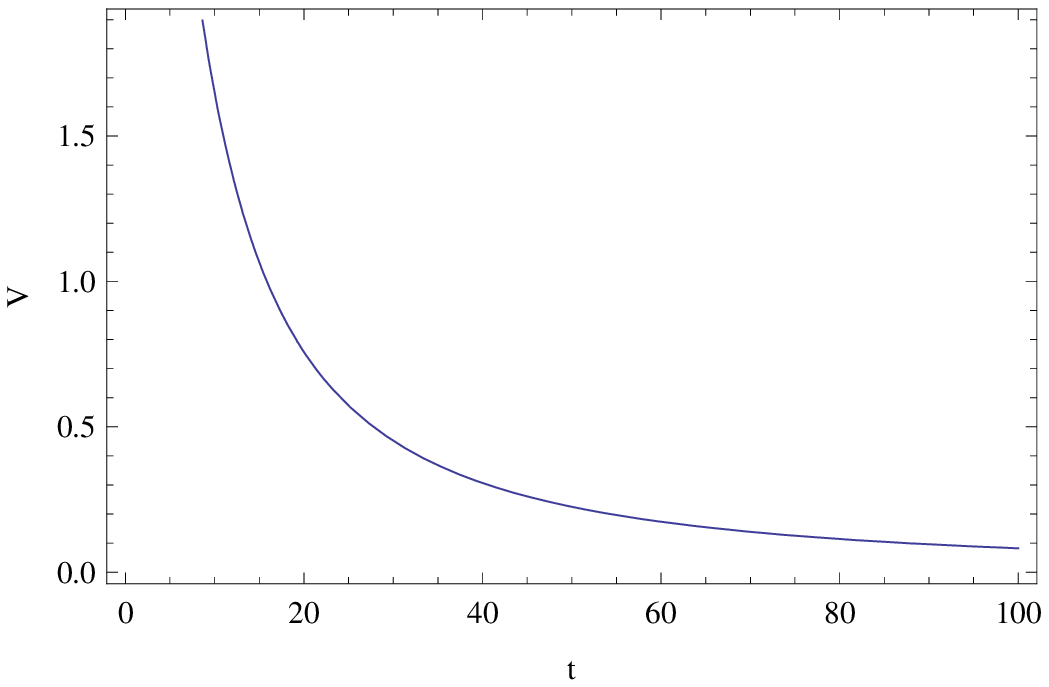}\\
\vspace{2mm}
~~~~~~~Fig.16~~~~~~~~~~~~~~~~~~~~~~~~~~~~~~~~~~~~~~~~~~~~~~~~~~~~~~~~~~~~Fig.17\\
\vspace{6mm}
\includegraphics[scale=0.7]{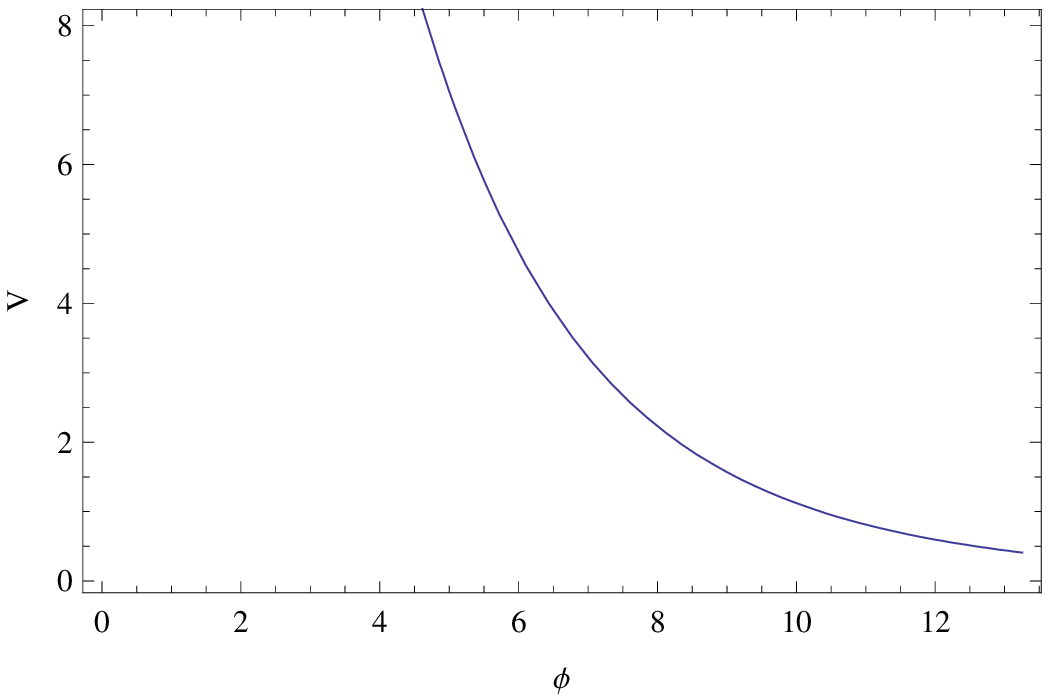}
\vspace{2mm}

Fig.18

\vspace{6mm} Figs. 16 - 17 show the variations of  $\phi$ and $V$
against $t$ and fig. 18  shows the variations of $V$ against
$\phi$, for $A = 1.1, B = 1.2, \lambda_{1} = 2, \lambda_{2} = 3, k
= 1, w = 1/3, \rho_{0}= 1$ in presence of phantom field in
logamediate scenario.
\vspace{6mm}
\end{figure}

\subsection{\bf\large{Logamediate Scenario}}

Consider a particular form of logamediate scenario, where the form
of the scale factors $a(t)$ and $b(t)$ are defined as [23]

\begin{equation}
a(t)=\exp (A(\ln t)^{\lambda_{1}})  \quad and  \quad  b(t)=\exp
(B(\ln t)^{\lambda_{2}})
\end{equation}
Using these expressions we have the Hubble parameter and it's
derivatives as,
\begin{equation}
H=\frac{A \ln t^{\lambda_{1}}\lambda_{1}+2B \ln
t^{\lambda_{2}}\lambda_{2}}{3t\ln t}, \quad \dot{H}=\frac{A \ln
t^{\lambda_{1}}\lambda_{1}(-1-\ln t +\lambda_{1})+2B \ln
t^{\lambda_{2}}\lambda_{2}(-1-\ln t +\lambda_{2})}{3 (t \ln
t)^{2}}
\end{equation}

and

\begin{equation}
\ddot{H}=\frac{A \ln t^{\lambda_{1}}\lambda_{1}(2+\ln t (3+ 2 \ln
t)-3(1+ \ln t)\lambda_{1}+\lambda_{1}^{2})+ 2 B \ln
t^{\lambda_{2}}\lambda_{2}(2+\ln t (3+ 2 \ln t)-3(1+ \ln
t)\lambda_{2}+\lambda_{2}^{2})}{3 (t \ln t)^{3}}
\end{equation}

 Hence, for a expanding Universe $A \lambda_{1} > 0$ and
$B \lambda_{2} > 0$. Also, from the derivative of Hubble parameter
we have $\lambda_{1}> 1 $ and  $\lambda_{2} > 1$  or if
$\lambda_{1} = \lambda_{1} = 1$,  then $A > 1$ and $B > 1$.\\

Now putting the assumed values of the scale factors in (10) and
(11), we have the required expressions for tachyonic field $\phi$
and its potential V as

\begin{eqnarray*}
\phi=\int\left[[\frac{\frac{2}{3}ke^{-2B (ln
t)^{\lambda_{2}}}+\frac{2}{3 t^{2} (ln t)^{2}}(-A (ln
t)^{\lambda_{1}})(1+A (ln t)^{\lambda_{1}})(\lambda_{1})^{2}+A (ln
t)^{\lambda_{1}}\lambda_{1}(1+ln t+2B (ln
t)^{\lambda_{2}})\lambda_{2})}{{\epsilon(e^{-2B (ln
t)^{\lambda_{2}}}k+\frac{2 A B (ln
t)^{-2+\lambda_{1}+\lambda_{2}}\lambda_{1}\lambda_{2}}{t^{2}}+\frac{B^{2}
(ln t)^{-2+2\lambda_{2}}\lambda_{2}^{2}}{t^{2}}-(e^{A (ln
t)^{\lambda_{1}}+2B (\ln
t)^{\lambda_{2}}})^{(-1-w)}\rho_{0})}}\right.
\end{eqnarray*}
\begin{equation}
\left.\frac{-B (ln t)^{\lambda_{2}}\lambda_{2}(-2(1+ ln t)+(2+B
(ln t)^{\lambda_{2}})\lambda_{2}))+(e^{A (ln t)^{\lambda_{1}}}+2B
(ln t)^{\lambda_{2}})^{(-1-w)}(1+w)\rho_{0}]}{\epsilon(e^{-2B (ln
t)^{\lambda_{2}}}k+\frac{2 A B (ln
t)^{-2+\lambda_{1}+\lambda_{2}}\lambda_{1}\lambda_{2}}{t^{2}}+\frac{B^{2}
(ln t)^{-2+2\lambda_{2}}\lambda_{2}^{2}}{t^{2}}-(e^{A (ln
t)^{\lambda_{1}}+2B (\ln
t)^{\lambda_{2}}})^{(-1-w)}\rho_{0})}\right]^{1/2} dt
\end{equation}
and
\begin{eqnarray*}
V(\phi)=e^{-2B (ln t)^{\lambda_{2}}}k +\frac{2A B (ln
t)^{-2+\lambda_{1}+\lambda_{2}}\lambda_{1}\lambda_{2}+B^{2}(ln
t)^{-2+2 \lambda_{2}}\lambda_{2}^{2}}{t^{2}}-(e^{A (ln
t)^{\lambda_{1}}+2B (ln t)^{\lambda_{2}}})^{-1-\omega}\rho_{0}
\end{eqnarray*}
\begin{eqnarray*} \left[[1-\frac{\frac{2}{3}ke^{-2B (ln
t)^{\lambda_{2}}}+\frac{2}{3 t^{2} (ln t)^{2}}(-A (ln
t)^{\lambda_{1}})(1+A (ln t)^{\lambda_{1}})(\lambda_{1})^{2}+A (ln
t)^{\lambda_{1}}\lambda_{1}(1+ln t+2B (ln
t)^{\lambda_{2}})\lambda_{2})}{{(e^{-2B (ln
t)^{\lambda_{2}}}k+\frac{2 A B (ln
t)^{-2+\lambda_{1}+\lambda_{2}}\lambda_{1}\lambda_{2}}{t^{2}}+\frac{B^{2}
(ln t)^{-2+2\lambda_{2}}\lambda_{2}^{2}}{t^{2}}-(e^{A (ln
t)^{\lambda_{1}}+2B (\ln
t)^{\lambda_{2}}})^{(-1-w)}\rho_{0})}}\right.
\end{eqnarray*}
\begin{equation}
\left.\frac{-B (ln t)^{\lambda_{2}}\lambda_{2}(-2(1+ ln t)+(2+B
(ln t)^{\lambda_{2}})\lambda_{2}))+(e^{A (ln t)^{\lambda_{1}}}+2B
(ln t)^{\lambda_{2}})^{(-1-w)}(1+w)\rho_{0}]}{(e^{-2B (ln
t)^{\lambda_{2}}}k+\frac{2 A B (ln
t)^{-2+\lambda_{1}+\lambda_{2}}\lambda_{1}\lambda_{2}}{t^{2}}+\frac{B^{2}
(ln t)^{-2+2\lambda_{2}}\lambda_{2}^{2}}{t^{2}}-(e^{A (ln
t)^{\lambda_{1}}+2B (\ln
t)^{\lambda_{2}}})^{(-1-w)}\rho_{0})}\right]^{1/2}
\end{equation}

Now putting the assumed values of the scale factors in (14) and
(15), we have the required expressions for phantom field $\phi$
and its potential V as

\begin{eqnarray*}
\phi=\int\left[-\frac{2}{3}ke^{-2B (ln
t)^{\lambda_{2}}}+2\frac{(-A (ln t)^{\lambda_{1}})(1+A (ln
t)^{\lambda_{1}})(\lambda_{1})^{2}+A (ln
t)^{\lambda_{1}}\lambda_{1}(1+ln t+2B (ln
t)^{\lambda_{2}})\lambda_{2})+}{3 t^{2}(ln t)^{2}}\right.
\end{eqnarray*}

\begin{equation}
\left.+\frac{B (ln t)^{\lambda_{2}}\lambda_{2}(-2(1+ ln t)(2+B
(ln t)^{\lambda_{2}})\lambda_{2}))}{3 t^{2}(ln t)^{2}}+(e^{A (ln
t)^{\lambda_{1}}+2B (\ln
t)^{\lambda_{2}}})^{(-1-w)}\rho_{0})\right]^{1/2} dt
\end{equation}
and
\begin{eqnarray*}
V=\frac{(A (ln t)^{\lambda_{1}} \lambda_{1}+2B (ln
t)^{\lambda_{2}}\lambda_{2})^{2}+A(ln
t)^{\lambda_{1}}(-1+\lambda_{1})\lambda_{1}+2B (ln
t)^{\lambda_{2}}(-1+\lambda_{2})\lambda_{2}}{3 t^{2}(ln t)^{2}}
\end{eqnarray*}
\begin{equation}
+\frac{1}{2}(e^{A(ln t)^{\lambda_{1}}+2B(ln
t)^{\lambda_{2}}})^{-1-w} (-1+w)\rho_{0}+\frac{2}{3}e^{-2B(ln
t)^{\lambda_{2}}}k
\end{equation}

And the dark energy density and mass of the universe for the two
cases are given by
\begin{equation}
\rho_{\phi}=e^{-2B (ln t)^{\lambda_{2}}}k +\frac{2A B (ln
t)^{-2+\lambda_{1}+\lambda_{2}}\lambda_{1}\lambda_{2}+B^{2}(ln
t)^{-2+2 \lambda_{2}}\lambda_{2}^{2}}{t^{2}}-(e^{A (ln
t)^{\lambda_{1}}+2B (ln t)^{\lambda_{2}}})^{-1-\omega}\rho_{0}
\end{equation}

\begin{equation}
Mass=\frac{e^{A (ln t)^{\lambda_{1}}}\left(k t^{2}(ln t)^{2}+B
e^{2 B (ln t)^{\lambda_{2}}}(ln t)^{\lambda_{2}}\lambda_{2}(2A (ln
t)^{\lambda_{1}}\lambda_{1}+B(ln
t)^{\lambda_{2}}\lambda_{2})\right)}{t^{2} (ln t)^{2}}
\end{equation}

From above we see that the expressions of tachyonic field and
phantom field and their corresponding potentials are very
complicated. The normal tachyonic field ($\epsilon=+1$) and
corresponding potential against time $t$ have been drawn in
figures 10, 11 respectively and the potential against the
corresponding field have been drawn in figures 12. Also the
phantom tachyon field ($\epsilon=-1$) and phantom field with the
corresponding potentials have been drawn in figures 13, 14, 16 and
17 respectively and the fields potentials against the
corresponding fields have been drawn in figures 15 and 18
respectively in logamediate scenario for $A = 1.1, B = 1.2,
\lambda_{1} = 2, \lambda_{2} = 3, k = 1, w = 1/3, \rho_{0}= 1$.
From figures 10-12, we see that the normal tachyonic field and
potential always increase with time and potential increases
against normal tachyonic field. From figures 13-15, we see that
the phantom tachyonic field and potential always decrease with
time and potential decreases with phantom tachyonic field. Also,
from figures 16-18, we see that the phantom field always increases
with time and potential always decreases with time and phantom field.\\

Now, we have graphically analyze the dark energy density for
phantom and tachyon scalar field in intermediate and logamediate
scenarios.

\begin{figure}
\includegraphics[scale=0.5]{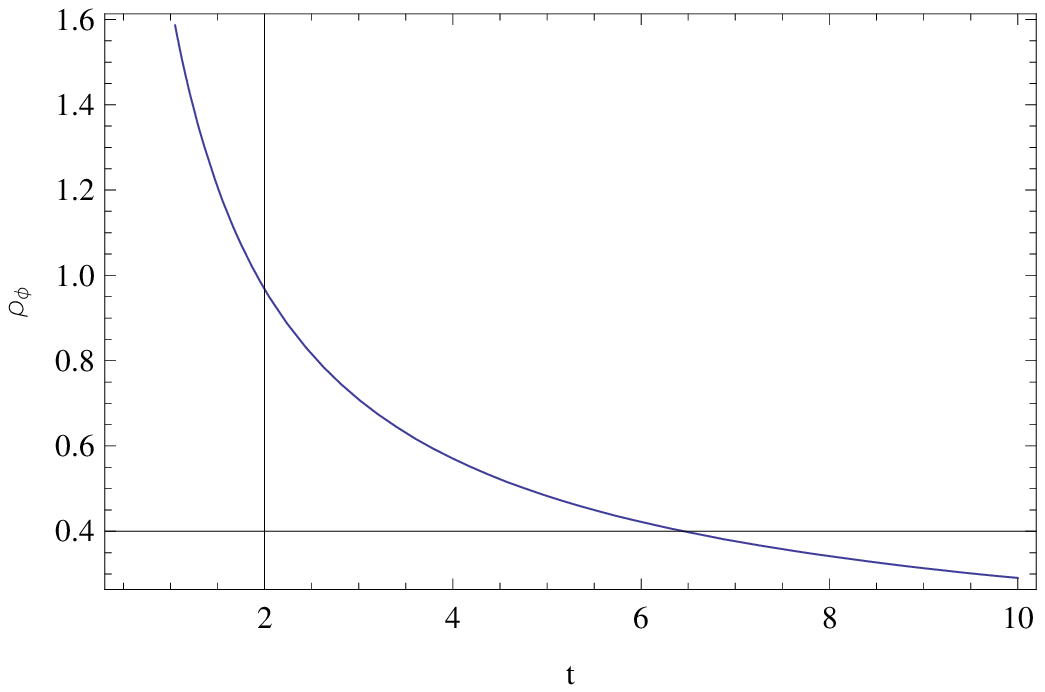}~~~~~~~~~~~~~~~
\includegraphics[scale=0.5]{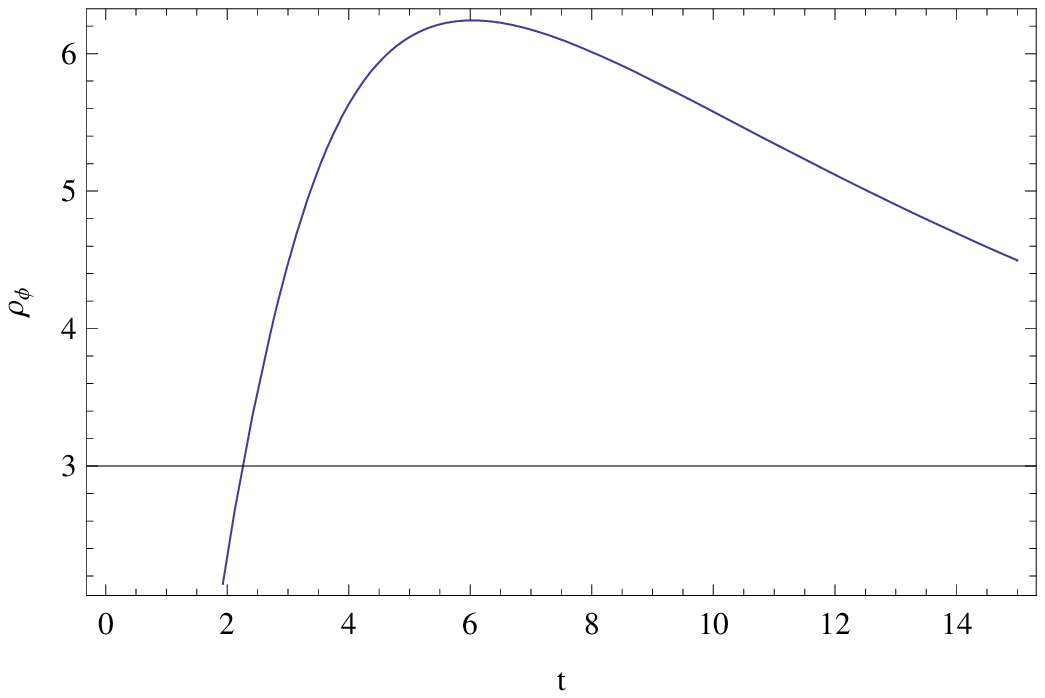}\\
\vspace{2mm}
~~~~~~~~Fig.19~~~~~~~~~~~~~~~~~~~~~~~~~~~~~~~~~~~~~~~~~~~~~~~~~~~~~~~~~~~~~~~Fig.20\\

\vspace{6mm}

Fig. 19 and Fig. 20 show the variations of $\rho_{\phi}$ (dark
energy density) against $t$ for Intermediate scenario and
Logamediate scenario . \vspace{6mm}
\end{figure}

\section{\normalsize\bf{Discussions}}

In this work, we have analyzed two scenarios namely,
``intermediate'' and ``logamadiate'' scenarios for closed, open
and flat anisotropic universe in presence of phantom field, normal
tachyonic field and phantom tachyonic field. We have assumed that
there is no interaction between the above mentioned dark energy
and dark matter. In these two types of the scenarios of the
universe, the nature of the scalar fields and corresponding
potentials have been investigated. In case of intermediate
scenario we see from figures 1 and 2 that the normal tachyonic
field $\phi$ increases and the potential $V(\phi)$ decreases with
time but remain positive. From the figure 3 it is clear that the
potential also decreases with the increase of the field. Where as
for phantom tachyon we see from the figures 4 and 5 that the field
and the potential both are increasing with time. Also from the
figure 6 we came to know that potential also increases with the
increase of the field. Also for phantom field we see from the
figures 7 and 8 that the field and the potential both are
increasing with time and from the figure 9 we came to know that
potential also increases with the increase of the field. For
intermediate scenario we express the dark energy density and mass
of the universe in term of cosmic time $t$ and from fig 19 we
came to know that the density is decreasing by the evolution of
the universe.\\

In case of logamediate scenario we see from figures 10 and 11 that
the normal tachyonic field $\phi$ increases and the potential
$V(\phi)$ increases with time but remain positive. From the figure
12 it is clear that the potential also increases with the increase
of the field. Where as for phantom tachyon we see from the figures
13 and 14 that the field and the potential both are decreasing
with time. Also from the figure 15 we came to know that potential
also decreases with the increase of the field. Also for phantom
field we see from the figures 16 and 17 that the field increases
and the potential decreases with time and from the figure 18 we
came to know that potential also decreases with the increase of
the field. For logamediate scenario we express the dark energy
density and mass of the universe in term of cosmic time $t$ and
from fig 20 we came to know that the density is gradually
decreasing by the evolution of the universe.\\

\textbf{References: }\\\\\
[1] A. G. Riess etal, \textit{Astron. J.} \textbf{116} 1009
(1988); S. Perlmutter etal, \textit{Astrophys. J.} \textbf{517}
565 (1999).\\\
[2] K. Abazajian etal, \textit{Astron. J.} \textbf{128} 502
(2004); Abazajian etal, \textit{Astron. J.} \textbf{129} 1755 (2005).\\\
[3] D. N. Spergel etal, \textit{Astrophys. J. Suppl.} \textbf{148} 175 (2003); {\it Astrophys. J. Suppl.} {\bf 170} 377 (2007).\\\
[4] P. J. E. Peebles and B. Ratra, \textit{Astrophys. J.}
\textbf{325} L17 (1988); R. R. Caldwell, R. Dave and P. J. Steinhardt, \textit{Phys. Rev. Lett.} \textbf{80} 1582 (1998).\\\
[5] C. Armendariz - Picon, V. F. Mukhanov and P. J. Steinhardt, \textit{Phys. Rev. Lett.} \textbf{85} 4438 (2000).\\\
[6] A. Sen, \textit{JHEP} \textbf{0207} 065 (2002).\\\
[7] R. R. Caldwell, \textit{Phys. Lett. B} \textbf{545} 23 (2002).\\\
[8] N. Arkani-Hamed, H. C. Cheng, M. A. Luty and S. Mukohyama, \textit{JHEP} \textbf{0405} 074 (2004).\\\
[9] F. Piazza and S. Tsujikawa, \textit{JCAP} \textbf{0407} 004 (2004).\\\
[10] B. Feng, X. L. Wang and X. M. Zhang, \textit{Phys. Lett. B}
\textbf{607} 35 (2005); Z. K. Guo, Y. S. Piao, X. M. Zhang and Y.
Z. Zhang, \textit{Phys. Lett. B} \textbf{608} 177 (2005).\\\
[11] L. Amendola, \textit{Phys. Rev. D} \textbf{62} 043511 (2000); X. Zhang, \textit{Mod. Phys. Lett. A} \textbf{20} 2575 (2005).\\\
[12] V. Sahni and Y. Shtanov, \textit{JCAP} \textbf{0311} 014 (2003).\\\
[13] A. Y. Kamenshchik, U. Moschella and V. Pasquier, \textit{Phys. Lett. B} \textbf{511} 265 (2001).\\\
[14] B. Chang, H. Liu, L. Xu and C. Zhang, {\it Chin. Phys. Lett.} {\bf 24} 2153 (2007).\\\
[15] S. -G. Shi,Y. -S. Piao and C. -F. Qiao, {\it JCAP} {\bf 0904} 027 (2009).\\\
[16] H. B. Benaoum, {\it hep-th}/0205140v1.\\\
[17] M. Sami, {\it Pramana} {\bf 62} 765 (2004).\\\
[18] U. Debnath, \textit{Class. Quantum Grav.} \textbf{25} 205019 (2008).\\\
[19] M. R. Setare, {\it Phys. Lett. B} {\bf 653} 116 (2007).\\\
[20] J. D. Barrow and N. J. Nunes \textit{Phys. Rev. D}, \textbf{76}, 043501 (2007).\\\
[21] J. D. Barrow, {\it Phys. Lett. B} {\bf 235} 40 (1990); J. D.
Barrow and P. Saich, {\it Phys. Lett. B} {\bf 249} 406 (1990); J.
D. Barrow, A. R. Liddle, and C. Pahud, {\it Phys. Rev.
D} {\bf 74} 127305 (2006).\\\
[22] K. S. Thorne, {\it Astrophys. J.} {\bf 148} 51 (1967).\\\
[23] S. Chakraborty and U. Debnath, {\it Int. J. Mod. Phys. A}
{\bf 25} 4691 (2010).\\\
[24]M. Jamil, F. M. Mahomed and D. Momeni, {\it Phys. Lett. B} {\bf 702} 315 (2011).\\\
[25]M. Jamil and A. Sheykhi, {\it IJTP} {\bf 50} 625 (2011).\\\
[26]M. U. Farooq, M. A. Rashid and M. Jamil, {\it IJTP} {\bf 49} 2278 (2010).\\\
[27]M. U. Farooq, M. Jamil and U. Debnath, {\it Astroph. Space Sci.} {\bf 334} 243 (2011).\\\
[28]M. Jamil and M. Akbar, {\it Gen. Rel. Grav.} {\bf 43} 1061 (2011).\\\
[29]M. Jamil and I. Hussain, {\it IJTP} {\bf 50} 465 (2011).\\\
[30]P. B. Khatua and U. Debnath, {\it IJMPA} {\bf 25} 4691 (2010).\\\
[31]R. Herrera and N. Videla, {\it Eur. Phys. J. C} {\bf 67} 499 (2010).\\\
[32]A. Cid and S. d. Campo, {\it JCAP} {\bf 1001} 013 (2011).\\\
[33]S. d. Campo, R. Herrera and A Toloza {\it Phys. Rev. D} {\bf 79} 083507 (2009).\\\
[34]U. Debnath, S. Chattopadhayay and M. jamil, \textit{arXiv:1107.0541v1 [physics.gen-ph]}  (2011).\\\

\end{document}